# Nonlocal quantum differentiation between polarization objects using entanglement

Vira R. Besaga,[1,*] Luosha Zhang,[2,a)] Andres Vega,[1] Purujit Singh Chauhan,[1, b)] Thomas Siefke,[1, b)] Fabian Steinlechner,[3] Thomas Pertsch,[1, b)] Andrey A. Sukhorukov,[4] and Frank Setzpfandt[1, b)]

[1] Friedrich Schiller University Jena, Institute of Applied Physics, Abbe Center of Photonics, Jena, 07745, Germany
[2] Institute of Microelectronics of the Chinese Academy of Sciences, 100029 Beijing, China
[3] Fraunhofer Institute for Applied Optics and Precision Engineering IOF, Jena, 07745, Germany
[4] ARC Centre of Excellence for Transformative Meta-Optical Systems (TMOS), Department of Electronic Materials Engineering, Research School of Physics, The Australian National University, Canberra, ACT 2600, Australia

[a)] Also at Friedrich Schiller University Jena, Institute of Applied Physics, Abbe Center of Photonics, Jena, 07745, Germany and at University of the Chinese Academy of Sciences, 101408 Beijing, China
[b)] Also at Fraunhofer Institute for Applied Optics and Precision Engineering IOF, Jena, 07745, Germany

*vira.besaga@uni-jena.de

For a wide range of applications a fast, non-destructive, remote, and sensitive identification of samples with predefined characteristics is preferred instead of their full characterization. Here, we report on the experimental implementation of a nonlocal quantum measurement scheme enabling to distinguish different transparent and birefringent samples by means of polarization-entangled photon pairs and remote state preparation. On an example set of more than 80 objects with varying Mueller matrices we show that only two coincidence measurements are already sufficient for successful discrimination in contrast to at least 8 required for a comprehensive inspection. The decreased number of measurements and the sample set significantly exceeding a typical set size for various problems demonstrate the high potential of the method for applications aiming at biomedical diagnostics, remote sensing, and other classification/detection tasks.

## 1    Introduction

For a broad range of fundamental and applied research areas the polarization of light serves either as a probe or a subject under study. Understanding the polarization response of an object of interest can reveal characteristic features in the sample's structure and/or behaviour, e.g., the birefringence of crystals and bio-tissues [1,2], polarization selectivity in reflection/transmission of periodic structures [3-5], and chemical composition via detection of chiral proteins, molecules, or their assemblies [6,7]. The knowledge about the object under study that can be thus acquired is then widely employed for optical biomedical diagnostics [8-10], the control of nonlinear phenomena [11], fundamental studies [12], technical characterization [13,14], remote sensing [15], as well as for currently expanding quantum technologies in the fields of communication, computing, and metrology [16,17].

The well-acknowledged techniques of classical optics for polarization analysis like Stokes and Mueller polarimetry [18] are essentially based on projective analysis of the polarization state or its change. A comprehensive characterization of a specimen implies its illumination with different polarization states, while the resulting intensity modulation for several polarization projections is detected. Here, the application of non-classical states of light rapidly gains interest [19,20] and particular attention is given to the possibility of nonlocal, or ghost, polarimetry, which has been already demonstrated in several configurations using classical and quantum correlations [21-25].

Recently, we have developed a theoretical model aiming at coincidence-based discrimination of polarization objects, where no polarization-sensitive detection after the sample is required and polarization-entangled photon pairs are employed to realize the measurement [26]. This measurement approach is schematically depicted in Fig. 1. Here, the signal photon from the entangled pair probes the sample and then undergoes a particular but fixed polarization transformation $P_p$. No polarization-resolving measurement takes place in this channel. Instead, the projective analysis is performed at the entangled idler photon in a different optical path, which does not interact with the sample. This is enabled via the concept of remote state preparation [27]. The relative coincidence counts between the signal and idler photons are measured for several projections $P_r$ of the idler photon, which serve as a response of the sample. The measured coincidence counts can be used as coordinates and define a virtual space where the polarization response of the object is distinctively mapped. The location of the sample in this measurement space is linked to its polarization properties, where the dimensionality of the virtual space depends on the number of projections $P_r$ used in the measurements. A reduction of the total number of required measurements for the projective analysis is achieved by an optimized selection of the polarization projections $P_r$ in the idler arm and of the single fixed polarization transformation in the signal arm [26].

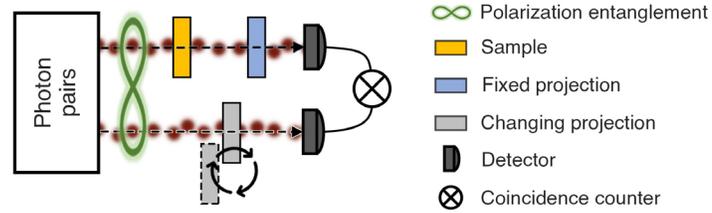

Fig. 1: Concept of the nonlocal quantum differentiation between objects using polarization-entangled photon pairs. Only one polarization projection is used in the probing arm and is kept fixed for all samples in the set to be discerned. A few alternating projections in a different optical path serve for polarization analysis. Relative coincidence counts between the probing and reference channels serve as a distinctive response of each sample.

In this letter, we report on experimental implementation of this non-destructive, remote, and sensitive polarization-based evaluation approach. Practically relevant samples often only slightly differ in the orientation of the polarization-dependent transmission axis or, in case of birefringent samples, orientation of the optical axis. We mimic a set of corresponding test samples with varying Mueller matrices by using different orientations of a linear polarizer and a quarter-wave plate. Using this set of objects as an example we show that only two projections $P_r$ and the corresponding coincidence measurements are already sufficient for successful differentiation between more than 80 sample types with differences in the orientation of the optical axis down to just 2 degrees. Hence, in our experiment the needed number of measurements is reduced multiple times in comparison to 8 or 16 measurements, which are required in the Jones representation [22] or the Mueller calculus [18], respectively, for full characterization of the sample's polarization properties. With this, the demonstrated approach of remote polarization-based detection becomes attractive for practical employment for, e.g., biomedical diagnostics or remote sensing with demand for real-time decision making.

## 2 Experimental realization

The experimental implementation of the nonlocal polarization-based identification of objects is sketched in Fig. 2 and follows the concept described above [26].

A source of entangled photon pairs is realized by two periodically-poled KTP (KTiOPO$_4$) crystals in type-II phase matching configuration, which are arranged with orthogonal optical axes within a polarization Mach-Zehnder interferometer [28]. The wavelength-degenerate photons are generated at 810 nm via the process of spontaneous parametric down-conversion (SPDC) and are spatially separated into two channels. Longpass filters ensure that no pump light enters the measurement arrangement. A polarization control module (PCM) consisting of two quarter-wave plates (QWP) enclosing a half-wave plate (HWP) is used to compensate for the minor temporal walk-off between the photons mainly related to the ambient conditions. The source nominally generates Bell state in the form $|\Psi^+\rangle = \frac{1}{\sqrt{2}}(|H\rangle|V\rangle + |V\rangle|H\rangle)$. The measured density matrix of the experimentally realized state is provided in the inset in Fig. 2. It is characterized by the linear

entropy of 0.13, concurrence of 0.88, and fidelity with the nominal Bell state of 0.90.

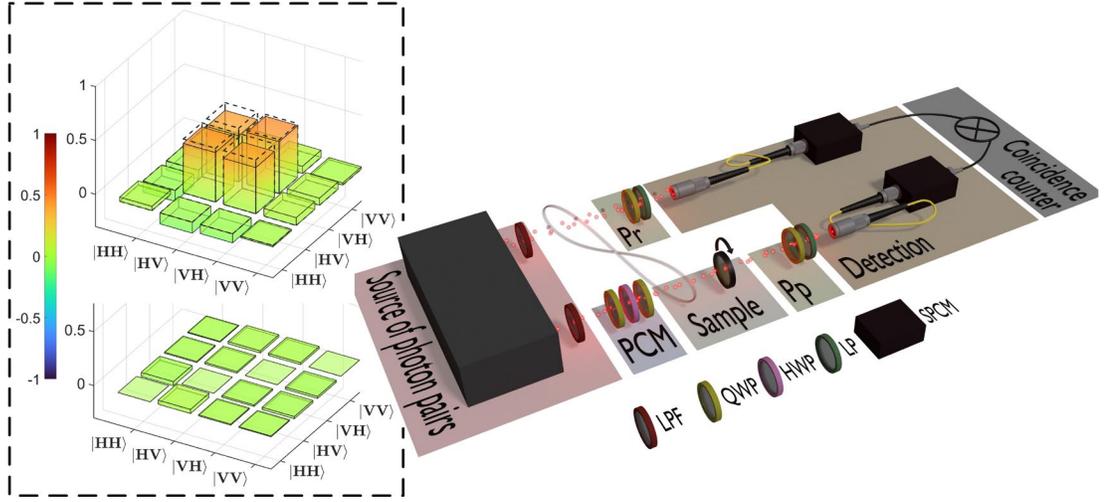

Fig. 2: Experimental setup used for the nonlocal differentiation of polarization objects. Inset shows the real (above) and imaginary (below) parts of a representative density matrix of the used two-photon state. Dashed black bars depict the ideal state of $1/\sqrt{2}(|H\rangle|V\rangle + |V\rangle|H\rangle)$ state. The colorbar represents the possible range of values of the density matrix elements. $H$ and $V$ in the axes tick labels mark horizontal $|H\rangle$ and vertical $|V\rangle$ polarization states. PCM: polarization control module; $P_p$: polarization projector in probe arm; $P_r$: polarization projector in reference arm; LPF: longpass filter; QWP: quarter-wave plate; HWP: half-wave plate; LP: linear polarizer; SPCM: single photon counting module.

The fixed polarization transformation $P_p$ in the signal arm and the projections $P_{r,n}$, $n = 1,2,3$, in the idler arm are implemented each as a combination of a QWP and a linear polarizer (LP). The latter is either a commercial LP of high extinction ratio or an in-house made partial polarizer with extinction ratio lower than 4:1 to generate partially polarized states. The transmitted photons are then detected with fiber-coupled single photon counting modules (SPCM) and the coincidences between the channels are counted for each projection combination in a sequential manner.

In order to ensure a well-predictable behavior of the objects serving as samples we utilized a conventional broadband linear polarizer (WP25M-VIS by Thorlabs) and an achromatic quarter-wave plate (AQWP05M-600 by Thorlabs) as the core elements. By rotating the LP or QWP in the optical path of the signal photon (see arrow in Fig. 2), we mimicked a substantial set of samples with varying Mueller matrices. Different orientations of LP and QWP have been obtained with an angular step of 1° within the range from 0° to 180°. Here, the orientation is identified by the angle between the transmission axis for LP and fast axis for QWP to the global vertical orientation in the optical arrangement. The samples obtained by rotating the LP will be in the following referred to as LP subset, and QWP subset for rotating the QWP, correspondingly.

In the experimental configuration described above, the signal photon undergoes polarization transformation $T_s = P_p M_s$ (i.e., interacts with the sample and passes the $P_p$ projector) and gets detected. Here, $M_s = R(\theta) M_{LP,QWP} R(-\theta)$ is sample's Muller matrix, $M_{LP,QWP}$ is the Mueller matrix of a vertically oriented LP or QWP, and $R(\theta)$ denotes rotation around the direction of light propagation by angle $\theta$. At the instance of signal photon detection, the idler photon, which does not interact with the sample, according to the concept of remote state preparation [27] can be written as the partial trace over the signal photon: $\rho_r = tr_P[(T_s \otimes 1)\rho_0(T_s \otimes 1)^\dagger]$, where $\dagger$ points the conjugate transpose. When the idler passes projection $P_{r,n}$ ($n$ = 1,2,3) and gets detected, the corresponding coincidence counts with the signal photon become $P_n = tr[(P_{r,n})\frac{\rho_r}{tr(\rho_r)}(P_{r,n})^\dagger]$ [26]. These expectation values $P_n$ ($n$ = 1,2,3) form the relative coordinates of the sample in a virtual n-dimensional space $\mathbb{P}$.

A representative measurement outcome is provided in Fig. 3 on example of $n$ = 1,2,3 (see Appendix

5.2 for details on $P_{r,n}$ realization). The coincidences $P_n$ between channels for each $P_{r,n}$ were counted for 1 s in a sequential manner for LP (Fig. 3a) and QWP (Fig. 3b) rotated with 1° step. The mean coincidence counts (solid curves) and 95% confidence intervals (shaded areas) are evaluated over 8 repetitions of the experimental run while Student's coefficient is employed to account for the limited number of experimental runs. To translate the measured counts to relative coordinates in the virtual $\mathbb{P}$ space, the measured data are scaled via dividing all the counts by the global maximal value in the dataset.

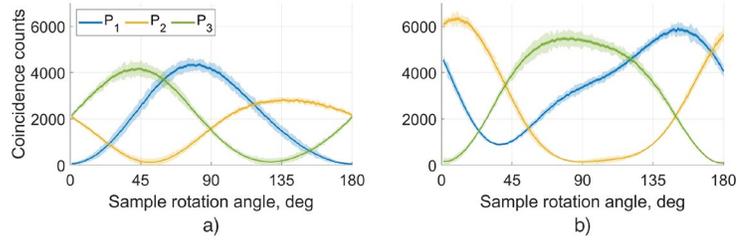

Fig. 3: Representative coincidences counted within 1 s intergration time between the signal and idler photons for each projection in the idler arm $P_{r1}$, $P_{r2}$, and $P_{r3}$ versus the angle of sample's orientation: a) for the linear polarizer and b) the quarter-wave plate in the path. The shown data has been obtained over at least 8 repetitions of the experiment. Solid lines depict the average coincidence counts. Shaded areas mark the 95% confidence interval. The latter is not limited to Poissonian detection statistics, but includes also the random experimental error and accounts for the limited number of experiment runs.

The samples' responses are then allocated in the n-dimensional $\mathbb{P}$ space using the obtained relative coordinates and their distribution serves for distinguishing the samples. In the following we provide the experimental results for the cases of three- (3D) and two-dimensional (2D) $\mathbb{P}$ space.

## 3 Results

### 3.1 Nonlocal differentiation with three coincidence measurements

Following our theoretical investigations [26] we first studied the feasibility of the experimental nonlocal differentiation between polarization objects using 3 coincidence measurements between the signal photon carrying the sample information and the idler photon analyzed at 3 different polarization projections, correspondingly. For this, we performed a set of experiments where LP or QWP with their principal optical axes oriented at different angles were placed into the signal arm of the experimental setup so that normal incidence of light was ensured and no extra illumination or imaging optics has been introduced into the optical path. The obtained experimental results are summarized in Fig. 4. We show here the virtual 3D space $\mathbb{P}$ defined by the coordinates $P_n$ ($n$ = 1,2,3) which correspond to the coincidences counted between the signal after manipulation $P_p$ and idler photons after the projections $P_{r,n}$. Each point marked as a dot (QWP subset) or cross (LP subset) depicts the mean value of the coincidence-based location obtained over at least 8 repetitions of the experiment. The shaded ellipsoids represent the 95% confidence regions whose semi-axes are equal to the 95% confidence intervals along the corresponding axes. Please refer to the inset of the figure for better visibility of ellipsoids of two grades of transparency (transparency coding explained further). The color of the shade represents the largest value of the standard deviation over three coordinates for a single sample over all repetitions of the experiment. The two loops formed by the ellipsoids correspond to the samples realized by QWP (outer loop, pointed out with a "QWP" label) and by LP (inner loop with "LP" label). For both subsets the filled black circles define the $0^{th}$ degree of orientation of the fast axis for the QWP and the transmission axis for the LP. As can be seen, the uncertainty regions for the objects with 1° rotation steps are overlapping. Hence, a reliable distinction between all these objects is not possible.

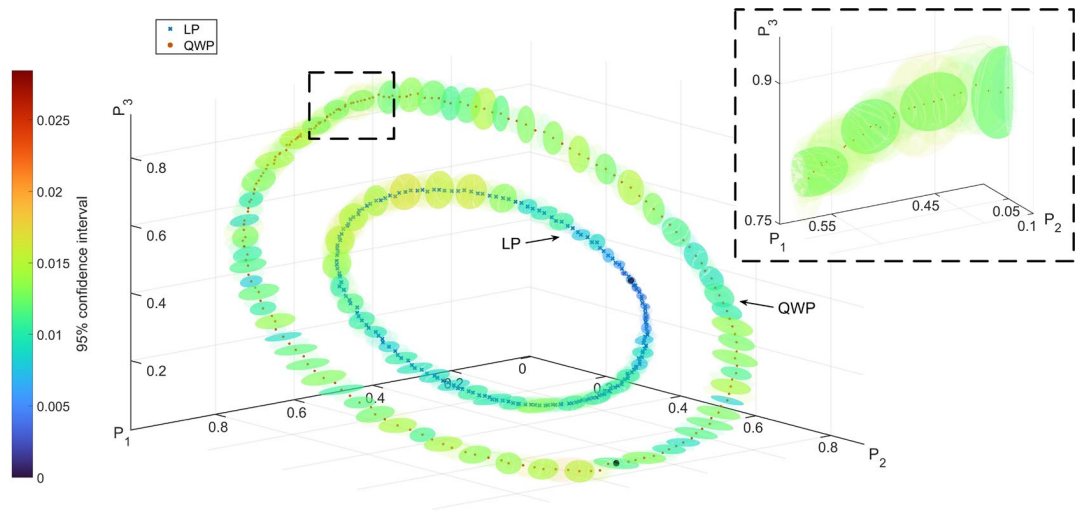

Fig. 4: Nonlocal differentiation between polarization samples with three coincidence measurements. $P_n$, $n = 1,2,3$, mark the relative coincidence counts for corresponding polarization projections $P_{r,n}$ used in the idler arm. Dots and crosses mark mean values of samples locations. The shaded ellipsoids show the 95% confidence regions. The color of the shade encodes the largest value of the standard deviation of the sample location over all coordinates. Less transparent ellipsoids mark distinguishable samples. Filled black circles define the $0^{th}$ degree of QWP (quarter-wave plate) and LP (linear polarizer) axis orientation. Inset zooms the domain defined by the black dashed rectangle.

However, a reduced by still sizeable subset of the measured objects can still be distinguished. The detection regions for these distinguishable objects are marked by the less transparent ellipsoids in Fig. 4. As can be seen from the figure, 42 samples in the LP subset and 60 samples in the QWP subset can be reliably separated from each other. However, since the volume and elongation of the uncertainty regions as well as the mean location points are unevenly distributed for the different angles of the LP and QWP subsets, the samples that can be distinguished are not evenly distributed as well. The median step in the optical axis orientation required to obtain separable samples is 4° for LP and 2° for QWP with corresponding maximal values reaching no more than 8° and 9°. For better visibility, one of the domains with a relatively large spread of the confidence regions (highlighted with the black dashed rectangle) is shown zoomed in the inset of the figure. At the same time, for a large range of angles the distinguishable samples have been obtained already with just 2° rotation difference for both LP and QWP.

### 3.2  Nonlocal differentiation with just two coincidence measurements

After proving experimentally the feasibility of three-projection nonlocal probing of polarization samples, we studied whether just two coincidence measurements can be sufficient for distinguishing the samples from the same set. The same LP and QWP have been used to generate a set of different samples in a similar manner as in previous experiment. Here, two of the previously employed polarization projections have been selected for the idler channel to define the 2D space for differentiation. The projection used for fixed polarization transformation in the signal arm has been adjusted to closer to an optimal one (see Appendix 5.2). For this, the conventional LP has been replaced with a custom in-house made LP of tailored extinction ratio. The corresponding results summarized over at least 8 measurement runs are represented in Fig. 5.

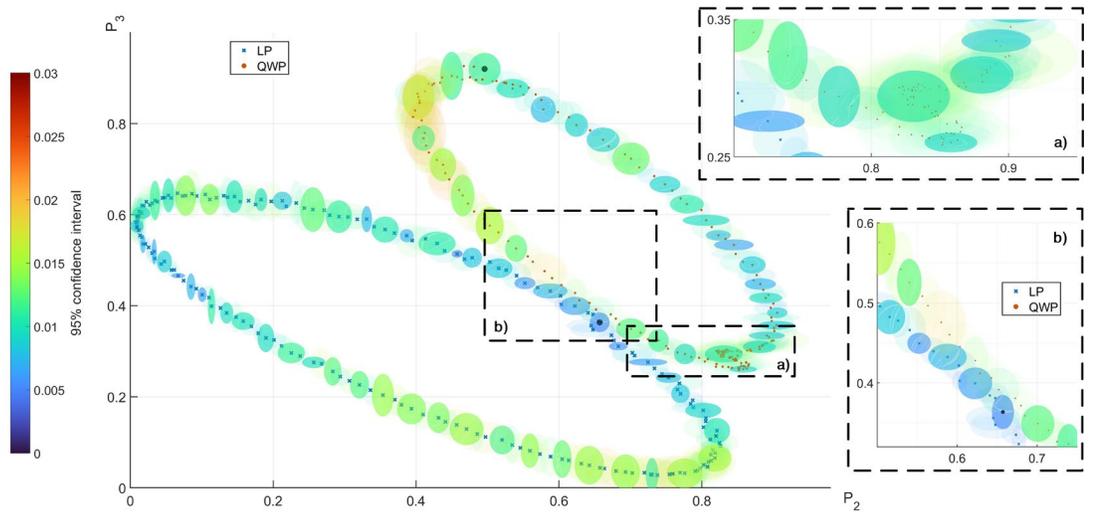

Fig. 5. Nonlocal differentiation between polarization samples with two coincidence measurements. $P_1$ and $P_2$ link the relative coincidence counts to the respective polarization projections in the idler arm $P_{r1}$ and $P_{r2}$. Dots and crosses mark mean locations of samples from QWP and LP subsets, correspondingly. The shaded ellipsoids show the 95% confidence regions. The color of the shade encodes the largest value of the standard deviation of the sample location over all coordinates. Less transparent ellipsoids mark distinguishable samples. Filled black circles define the $0^{th}$ degree of QWP and LP axis orientation. Insets a) and b) zoom the domains marked with black dashed rectangles and labeled respectively.

Analogous to the previous case, the figure shows the space $\mathbb{P}$, which now is just two-dimensional based on the two projections $[P_{r1}, P_{r2}]$ in the idler arm. Similar coding as in Fig. 4 is used for color and shading to represent the measurement statistics. Accounting for the 2D representation, the 95% confidence regions are depicted here as ellipses. Here as well, the mean values of the corresponding samples' locations are marked as crosses for LP and dots for QWP subsets, respectively.

The change of the polarization transformation implemented in the signal arm has resulted in a changed shape of the overall distribution of the samples in the virtual space. In particular, for the QWP the measured points are overlapping in a range of approx. from 60° to 110° of the fast axis orientation. Here, only three samples can be separated from each other. This region is highlighted with a black dashed rectangle labeled "a)" and shown zoomed in the corresponding inset. The overall number of distinguishable samples from the QWP subset with just two coincidence measurements is thus decreased to 34. The median step was found to be smaller than for the three-projection measurement and is equal to 3°. The same median step required to separate different samples has been detected for LP subset, where the distribution of measured responses was found similar to the previous case. The points are allocated along a single-loop curve with no self-crossing points and no exceptional regions, like those that appeared in case of QWP subset. With this, the full range of angles with 54 different samples could be separably projected in the 2D space of coincidence counts. For both LP and QWP subsets the maximum step in the rotation angle was defined as 8°, where the exceptional range from Fig. 5 inset "a)" was excluded. The minimal sufficient step is again 2°. However, the distribution of the sample responses is characterized by one more particular region which is highlighted with a black dashed rectangle labeled "b)" and shown zoomed in the respective inset. As one can see, the confidence regions of three samples from the QWP subset initially found as distinguishable, coincide with the confidence regions of the closely allocated samples from the LP subset. Although, the mean values of all these samples are still separable, three mentioned samples from the QWP set have been excluded from the list of distinguishable objects and thus their confidence regions have been encoded with higher transparency. Nevertheless, 84 samples in total can be reliably separated under the implemented experimental conditions which significantly exceeds a typical set size for various applications and shows the sufficiency of just two coincidence measurements for practical differentiation between polarization objects.

## 4  Discussion and conclusions

The presented results have revealed several points worth of particular attention.

The first one relates to the spread of the samples responses in $\mathbb{P}$ space defined by relative coincidence counts. The measured loops have been projected in the $\mathbb{P}$ space not in one plane and the resolution in terms of orientation angle could not be achieved better than 2 °. This can be explained by the combination of several factors. On the one hand side, the polarization transformations for idler arm have been realized using combinations of conventional QWP and LP which were close to optimal projections but still not most advantageous. This issue can be mitigated by the use of projections to partially polarized states. Using customized polarization manipulators that result in partially polarized states instead of the conventional LP and QWP used in our feasibility demonstration, the distance of the measured sample locations can be enlarged [26]. On the other hand side, the level of entanglement of the two-photon state employed in the experiment and decoherence occurring in the samples have to be accounted for as well. For our experiments the target polarization transformation was optimized considering the nominal Bell state in the form of $|\Psi^+\rangle = \frac{1}{\sqrt{2}}(|H\rangle|V\rangle + |V\rangle|H\rangle)$ (refer to Appendix 5.2 for details). At the same time, the concurrence of the utilized polarization-entangled photon pairs during the reported experiments has been observed at the level of 0.88 (for details on source characterization refer to Appendix 5.1). The latter is supposed to cause slight shrinkage of the loops of LP and QWP in terms of the reduced state of the idler photon and thus changed projections in the space of coincidences [26].

The second point to discuss is related to the experiment with two projections in the idler arm. As discussed above, a combination of conventional LP and QWP has been initially used in the signal arm as well (see Appendix 5.2). Although, being again not optimal, the resulting polarization transformation $P_p$ performed sufficiently well to enable differentiation between different samples with three coincidence measurements. It appeared, however, to be not suitable for the case of only two projections. To introduce the necessary partial polarization contribution we replaced the conventional LP with a custom linear polarizer of tailored extinction ratio (lower than 4 to 1), which has been manufactured in-house (see Appendix 5.3). This allowed us to distribute the sample responses over the 2D space of coincidence counts in a more advantageous manner. At the same time, this resulted in a slightly changed shape of the loop for samples within the QWP subset and the appearance of the exceptional region where fewer samples could be reliably separated (see Sec. 3.2 and Fig. 5 inset b). This issue can be addressed in the future with a custom phase retarder instead of the conventional QWP as well as further refinement of the numerical procedure for optimizing the polarization projectors in both signal and idler arms.

The advancements discussed above are expected to enable broader flexibility for realization of the optimal polarization transformations $P_p$ and $P_{r,n}$. This, in turn, would eliminate the partially overlapped and exceptional domains from insets in Fig. 5, but also would allow to experimentally perform differentiation between even larger sets of samples and/or samples with slighter changes in polarization behavior. At the same time, the results presented in this letter prove in practice the feasibility of the proposed method for nonlocal differentiation between objects with no polarization analysis after the sample and the reduced total number of the required measurements.

Besides this, the advantages of the suggested approach are expected to include the high sensitivity due to the inherently high signal-to-noise ratio of the coincidence-based detection and non-destructiveness in the context of light-induced damage of the sample. The latter is a straightforward result of the nonlocal, or ghost, configuration, but can be auxiliary engineered by using the non-degenerate polarization-entangled photon pairs, where the sample could be probed at the least harmful spectral range and the idler photon could be analyzed in the range where highly-efficient optics and detectors are available.

To sum up, in this letter, we reported on the first experimental demonstration of nonlocal differentiation between polarization objects using the polarization-entangled photon pairs, where no polarization analysis takes place after the sample. With just two polarization projections in the remote optical channel, single fixed polarization transformation in the local channel, and respective coincidence measurements between them, we demonstrated successful differentiation of more than 80 samples. While the proposed method substantially decreases the required number of measurements in comparison to full characterization, we were able to distinguish objects characterized by difference in the principal optical axis orientation down to just 2 degrees. Taking into account further advantages of nonlocal coincidence based probing, like non-destructiveness and low noise, there is a high potential for practical implementation of the approach, particularly for applications in biomedical diagnostics, remote sensing, and other classification/detection tasks with the demand for highly accurate real-time decision making.

## 5    Appendices

### 5.1    Characterization of the source of polarization-entangled photon pairs

Prior to every experiment on the nonlocal differentiation between samples, the photon state is characterized via the quantum state tomography [29]. For this, a combination of QWP and LP in each channel (in place of $P_p$ and $P_r$ manipulators in Fig. 2) is used to realize 16 polarization projection combinations in the $|H\rangle, |V\rangle$; $|D\rangle, |A\rangle$; and $|L\rangle, |R\rangle$ bases [30]. These notations stand for horizontal, vertical, diagonal (+45 °), anti-diagonal (-45 °), right-circular, and left-circular polarization states. The coincidences are counted using a time tagger with 1 ps temporal resolution within a 3 ns coincidence window over 10 s integration time. The raw coincidence counts are corrected for accidental counts, detector efficiency, and intensity drift [29]. The corrected values of coincidence counts are then used to reconstruct the density matrix of the state by the maximum likelihood method [30].

### 5.2    Realization of polarization projections in the probing and reference arms

An optimal polarization transformation in the sample arm is responsible for ensuring the distinct response of different samples in terms of polarization. When designing our experiments, we started off with numerical search for the optimal transformation $P_p$ considering the nominal state of our source of polarization-entangled photon pairs. Here, the objective function of the optimization process based on Matlab® *optimset* function was following our previous theoretical model [26] and was not restricted to feasibility of its implementation with conventional optical components. An example of such a versatile solution which could be realized in a form of a metasurface is represented in Table 1. Using this transformation as a reference, we searched for the closest experimentally feasible solutions which could be realized with different phase retarders and linear polarizers only. As a result, we were able to find the respective transformations when combinations of a QWP and either a commercial high-extinction ratio or custom tailored-extinction ratio LP were used. The corresponding Mueller matrices, which were realized in the experiments with three and two coincidence measurements, are represented in Table 1 as well. For this, the QWP was oriented with its fast axis at 62 ° and LP with its transmission axis at 90 ° to the global vertical orientation in the optical arrangement. Similar approach has been used when implementing polarization transformations for the idler photon. Overview of the orientation angles of the QWP and LP used in the experiments is provided in Table 2.

### 5.3    Manufacturing of a partial polarizer

The partial polarizer has been created on a fused silica substrate (Siegert) that is initially cleansed using piranha solution. Following this, 25 nm of chromium is added using ion beam sputter deposition (Ionfab 300LC (OIPT)). Next, OEBR-CAN038 AE 2.0CP (Tokyo Ohka Kogyo Co. LTD) EBL (electron beam lithography) resist is applied and patterned using a Vistec 350OS electron beam writer with character projection apertures [31]. The resultant grating pattern has a period of 300 nm which is transferred into the chromium layer through ion beam etching using the Ionfab 300LC (OIPT). Following this, the remaining resist is removed using oxygen plasma etching before the substrate is diced to size. The extinction ratio of the manufactured polarizer has been measured at 810 nm as 3.7 to 1 using a Perkin Elmer Lambda 950 spectrophotometer.

| | | | | |
|---|---|---|---|---|
| $P_p$ optimal | 0.6363 | -0.2086 | 0.1488 | -0.2716 |
| | -0.2086 | 0.3549 | 0.2452 | 0.3505 |
| | 0.1488 | 0.2452 | 0.1604 | -0.4491 |
| | 0.2716 | -0.3505 | 0.4491 | -0.1210 |
| $P_p$ for $\mathbb{P} = [P_1, P_2, P_3]$ | 0.5000 | -0.5000 | 0 | 0 |
| | -0.1563 | 0.1563 | 0 | 0 |
| | 0.2318 | -0.2318 | 0 | 0 |
| | 0.4145 | -0.4145 | 0 | 0 |
| $P_p$ for $\mathbb{P} = [P_1, P_2]$ | 0.6351 | -0.3649 | 0 | 0 |
| | -0.1141 | 0.1986 | -0.2410 | 0.4310 |
| | 0.1691 | -0.2944 | 0.3573 | 0.2907 |
| | 0.3025 | -0.5266 | -0.2907 | 0 |

Table 1. Mueller matrices (not normalized) of the fixed polarization transformation in the sample arm: optimal and the ones employed in the experiments with three and two coincidence measurements.

| | $P_p$ | $P_{r,1}$ | $P_{r,2}$ | $P_{r,3}$ |
|---|---|---|---|---|
| QWP | 62° | 170° | 18° | 45° |
| LP | 90° | 7.5° | 110° | 34° |

Table 2. Orientations of quarter-wave plates and linear polarizers (both conventional and custom) used for realizing polarization projections in the sample and reference arms within the reported experiments.

## Acknowledgements

The fabrication of the partial polarizer samples within this work was partly carried out by the microstructure technology team at IAP Jena. The authors would like to thank them for providing the fabrication facilities, carrying out processes and providing support.
This work has been funded by the German Ministry of Education and Research (вЂњQuantIm4LifeвЂќ project, FKZ 13N14877). V.B. thanks for funding of this work also to ProChance-career program (AZ 2.11.3-A1/2022-01) of the Friedrich Schiller University Jena. This project (20FUN02 POLight) has received funding from the EMPIR programme co-financed by the Participating States and from the European UnionвЂ™s Horizon 2020 research and innovation programme. A.A.S. also acknowledges support from the UA-DAAD exchange scheme and Australian Research Council (DP190101559, CE200100010).

## Author declarations

### Conflict of Interests

The authors declare no conflict of interest.

### Author contributions

V.R.B. designed and conducted the experiments, collected and analyzed the data, wrote the manuscript, L.Z. supported construction and alignment of the experimental arrangement, A.V. developed the used theoretical model and performed simulations, P.S.C. and F.St. designed and constructed the source of

polarization-entangled photon pairs, T.S. designed and manufactured the custom partial polarizer, wrote the manuscript, Fr.S. wrote the manuscript, A.A.S. and Fr.S. conceived the study, T.P., A.A.S., and Fr.S. supervised the research. All authors reviewed the manuscript.

## Data availability

The data that support the findings of this study are available from the corresponding author upon reasonable request.